\documentclass{article}
\usepackage{iclr2021_conference,times}
\usepackage{hyperref}
\usepackage{url}
\usepackage{graphicx}
\usepackage{listings}
\title{TikTok Engagement Traces Over Time and Health Risky Behaviors: Combining Data Linkage and Computational Methods}

\author{Xinyan Zhao\\
Hussman School of Journalism and Media\\
University of North Carolina-Chapel Hill\\
\And
Chau-Wai Wong\\
Electrical and Computer Engineering\\
NC State University}

\iclrfinalcopy 
\begin{document}

\maketitle

\begin{abstract}
Digital technologies and social algorithms are revolutionizing the media landscape, altering how we select and consume health information. Extending the selectivity paradigm with research on social media engagement, the convergence perspective, and algorithmic impact, this study investigates how individuals’ liked TikTok videos on various health-risk topics are associated with their vaping and drinking behaviors. Methodologically, we relied on data linkage to objectively measure selective engagement on social media, which involves combining survey self-reports with digital traces from TikTok interactions for the consented respondents (\textit{n} = 166). A computational analysis of 13,724 health-related videos liked by these respondents from 2020 to 2023 was conducted. Our findings indicate that users who initially liked drinking-related content on TikTok are inclined to favor more of such videos over time, with their likes on smoking, drinking, and fruit and vegetable videos influencing their self-reported vaping and drinking behaviors. Our study highlights the methodological value of combining digital traces, computational analysis, and self-reported data for a more objective examination of social media consumption and engagement, as well as a more ecologically valid understanding of social media’s behavioral impact.\vspace{2mm}
\end{abstract}

\section{Introduction}

Digital technologies and social
algorithms are revolutionizing the media landscape, altering how we select and
consume health information. Unlike the traditional mass media environment,
which offers limited choices of mediums and sources leading to more uniform
experiences across audience groups (Meyrowitz, 1986), social media present a
high-choice environment characterized by \textit{curated flows} (Thorson \&
Wells, 2015). Namely, social media users often find themselves at the core of
their personalized information environments (Zhao et al., 2022), embedded
within multiple, intersecting flows of content curated based on a variety of
factors including user preferences and behaviors, algorithmic inferences, social
contacts, and strategic communication.

The selectivity paradigm in media
effects and communication research emphasizes that individuals selectively
consume social media content based on personal preferences and are influenced
by the content they select (for a review, see Valkenburg, 2022). Extending the
selectivity paradigm to the high-choice information environment, the
convergence theory underscores individuals’ selective consumption of a range of
relevant content across key attributes of information systems, such as sources
or mediums (Anthony et al., 2013; Jenkins, 2006; Zhao et al., 2022). However,
extant work has focused on social media \textit{exposure} and how it is shaped by
algorithmic, individual, and social factors in the realm of political communication
(e.g., Guess et al., 2023; Thorson et al., 2021), seldomly exploring one’s \textit{engagement}
with \textit{a selective set} of social media content regarding health topics,
and their impact on personalized content flows on social media. It is crucial
to study \textit{selective engagement}, which highlights how users’ engagement
behaviors, such as likes and shares, interactively shape their subsequent media
exposure and engagement through a dynamic feedback loop where algorithms
continuously adjust the content shown to users based on their interaction patterns.
Additionally, given that a significant portion of the population worldwide uses
social media for health information (Jia et al., 2021), it is fundamental
to delve deeper than media exposure and examine the impact of social media engagement
on shaping personalized health-related content and its subsequent effects.

TikTok, with its 834 million global
users as of 2024, offers users a wealth of engagement opportunities through
immersive and interactive video experiences (Howarth, 2024). Despite its
popularity, especially among the younger population in the U.S., there has been
a dearth of research on individuals’ engagement with TikTok videos regarding
health content and its behavioral effects. Persuasion research indicates that
one processes information in relation to another (Akin et al., 2019),
suggesting that engagement with varied health-risk content can jointly affect
one’s behaviors. For example, consuming and interacting with TikTok videos
related to vaping and drinking, as well as those promoting fruit and vegetable
intake, may collectively shape the personalized information intake and
influence vaping and drinking behaviors. Therefore, this study investigates how
the liked TikTok videos regarding a variety of health-risk topics relate to vaping
and drinking behaviors.

Methodologically, research has shown
that accurately recalling media consumption poses a significant challenge for
survey respondents (Prior, 2009; Ohme, 2020; Scharkow,
2016). This has been intensified by the omnipresence of social media and the
fluid, often overlapping nature of online interactions, as well as the
proliferation of mobile devices (Araujo et al., 2017; Naab et al., 2018).
However, most research on social media consumption and engagement, particularly
in the health realm, relies on self-reports, potentially suffering from the
concern of measurement validity. This study addresses the limitation by
adopting a data linkage approach combining TikTok digital traces and self-reported
survey data (De Vreese et al., 2017; Stier et al., 2019; Otto et al., 2023). We
conducted an online survey and linked it with TikTok data by securing consent
from participants to access their historical TikTok activities through a
third-party application. This enabled us to download the TikTok videos that
respondents had liked over time and conduct a computational analysis. Through
data linkage, we associated the topics of videos liked by individuals at
different time points with their self-reported behaviors on vaping and
drinking. Our findings suggest that smoking, drinking, and fruit and vegetable
consumption videos individuals liked on TikTok related to their self-reported
vaping and drinking behaviors. 

\subsection{The Selectivity Paradigm to Social Media Consumption and Engagement}

The selectivity paradigm in media
effects and communication research theories emphasizes that individuals are
limited in the amount of media content they can focus on from the many
available, thus they choose the information based on personal dispositions,
needs, and desires, and are influenced only by the information they select (for
a review, see Valkenburg, 2022). This paradigm is well represented by selective
exposure theories (Zillmann \& Bryant, 1985), which assume that individuals
are motivated by cognitive and psychosocial factors to select and consume
certain media content. In the context of health messaging and behavioral
change, individuals can seek information on certain health topics to promote
health behavioral changes (Knobloch-Westerwick et
al., 2013). For instance, selective consumption of videos on drinking can reinforce
existing drinking behaviors (self-bolstering) or encourage an increase in
drinking behaviors (self-motivating). Knobloch-Westerwick
et al. (2017) found that more time spent viewing messages promoting health
behaviors results in a shift in attitudes toward the advocated health behaviors.
Despite the value of selective exposure theories in elucidating the mechanisms
of selective consumption in traditional media contexts, they do not fully
explicate or explain selective engagement in social media contexts. Distinct
from selective exposure, selective engagement behaviors can shape users’
personalized information environments and subsequent social media exposure
through a dynamic feedback loop. This loop is driven by algorithms that respond
to users’ engagement behaviors, such as likes and shares, by adjusting the
content they are shown in the future. Therefore, our work aims to extend the
selectivity paradigm by theorizing social media selective engagement based on the
emerging literature on social media convergence and algorithmic impact.

As social media increasingly blur
the lines between content production and consumption and merge the roles of
message senders and receivers, the high-choice social media environment adds to
the complexity and dynamics of individual information selection. To capture the
complexity, the \textit{convergence} perspective offers a theoretical explanation
of the personalized information environment by highlighting the simultaneous
consumption of relevant information across key attributes of information
systems, such as sources or mediums (Anthony et al., 2013; Jenkins, 2006; Zhao et
al., 2022). This perspective highlights how individuals navigate and integrate
a multitude of information, blending them into their daily media consumption to
create a unique, personalized media experience. According to Hermann et al.
(2023), social media users often find themselves in distinct, idiosyncratic
digital environments characterized by high levels of personalization, varying
significantly from one user to another and across platforms (e.g., TikTok, Facebook,
Instagram). This line of work underscores the need for an \textit{ecological and nuanced
}approach to studying selective engagement, focusing on how an individual
chooses to engage with a collection of content that forms their personalized
information environment on a particular platform. We take the challenge to
address this perspective, which remains largely unexplored in the literature.

Further, recent work on algorithmic impact has revealed
the significant role of individual interactive behaviors, in conjunction with algorithmic
and social factors, in shaping curated content flows on social media (Cho et
al., 2020; Guess et al., 2023; Pariser, 2011; Thorson et al., 2021; Wells \&
Thorson, 2015). According to Thorson et al. (2021), Facebook users can
intentionally tailor their information environment to their interests by
interactive behaviors such as liking specific content, signaling to platforms
like Facebook their interest in receiving similar content in the future.
However, it is unknown how selective engagement behaviors, coupled with
algorithmic impact, might amplify one’s engagement with certain kinds of
content over time. Additionally, existing research, largely focused on content exposure
on traditional platforms such as Facebook, concentrates on how political biases
drive personal content choices. This study expands the theoretical horizon of
the selectivity paradigm to investigate selective engagement with a collection
of health-related videos on TikTok and the behavioral outcomes.

\subsection{Selective Engagement Behaviors and Liked TikTok Videos}

Given social media’s impact on both
content exposure and individual engagement with that content, it is fundamental
for researchers to extend their focus beyond selective exposure and investigate
selective engagement behaviors on social media and its patterns over time. Engagement
is defined as the “willingness to invest in the undertaking of focal
interactions with particular engagement objects” (Hollebeek
et al., 2016, p. 2). Despite the multidimensional nature of social media
engagement and measurement nuances, this study relies on “likes” on TikTok as a
proxy of social media engagement behaviors (Hollebeek
et al., 2014; Zhao \& Chen, 2022), employing the behavioral traces to examine
the type and variety of health videos with which TikTok users engage over time
and their association with behavioral outcomes.

Rooted in prosocial desires and
emotional needs (Brandtzaeg \& Haugstveit,
2014), the interaction of liking allows users to share information, communicate
approval, and express a desire for connection and enjoyment (Brandtzaeg \& Haugstveit, 2014;
Khan, 2017). Through likes, users navigate and contribute to a complex web of
social interactions, from maintaining relationships to signaling agreement,
thereby enriching the interconnectedness of digital environments (Bucher \&
Helmond, 2018). Liking also significantly influences individuals’ future social
media exposure by aligning their exposure more closely with the content they
choose to engage with (Thorson et al., 2021). These interactions create a
feedback loop that personalizes the content experience, fostering prolonged
engagement and enabling advertisers to target content more effectively toward
specific demographics (Niland et al., 2016; Pokhrel et al., 2018). For example,
TikTok’s (2020) For You feed algorithm recommends content based on a
combination of factors, starting from preferences as a new user and adjusting
based on user interactions such as the videos one likes and video information,
to “ensure users see more of what they enjoy.” (p. 1) 

We choose to focus on the “liked”
videos by TikTok users over time, as the concept of behavioral engagement
aligns closely with the selectivity paradigm in which users choose to like
certain videos while ignoring others, reflecting how individuals exercise behavioral
choice in their interactions with social media content. Additionally, liking content
on social media demands less cognitive effort compared to more labor-intensive
engagement behaviors like sharing or commenting. This makes liking a more
extensively used form of engagement behaviors across platforms (i.e.,
“universal currency”; Bucher \& Helmond, 2018, p. 25). As we aim to
understand what types of health-related videos people interact with over time
and the downstream effects on behaviors, using “liking” as a proxy of
engagement behaviors allows us to capture a large and diverse dataset of health
content people interact with at different times. In the following section, we
discuss the data linkage approach to examine the outcomes of liked TikTok
videos of varied health topics.

\subsection{Data Linkage Approach to Studying Outcomes of Liked TikTok Videos}

Communication researchers have employed
two methodological paradigms to study media effects: surveys and computational
approaches (De Vreese et al., 2017; Ohme et al., 2023). While quantitative
social scientists have traditionally relied on surveys to study media use,
political participation, and health attitudes, the rise of digital technologies
has highlighted the imprecision and unreliability of self-reported data often
assessed retrospectively (Araujo et al., 2017; De Vreese \& Neijens, 2016; Parry et al., 2021; Scharkow,
2019). Participants struggle to accurately recall their media consumption (
Niederdeppe, 2016; Scharkow,
2016) and tend to overreport internet use in general (Araujo et al., 2017; 
Scharkow, 2016), cross-cutting exposure online (Song \&
Cho, 2021), as well as political activities on social media (
Haenschen, 2020). In contrast, computational methods enable
the collection of digital traces in a precise, nonintrusive way, offering
granular and nuanced insights into a variety of interaction behaviors. Despite
their precision, computational research primarily yields descriptive findings
limited to specific platform users, limiting the ability to infer the effects
or outcomes of analyzed content patterns.

Data linkage is a promising
methodology to integrate survey self-reports and digital traces, enhancing the
accuracy and ecological validity of media effect studies in social media
contexts (De Vreese et al., 2017; Stier et al., 2019; Otto et al., 2023). This
approach allows for the integration of survey responses and digital traces at
both aggregate and individual levels (Stier et al., 2019), utilizing
“user-centric” and “site-centric” strategies (De Vreese \& Neijens, 2016). On the one hand, aggregate-level linkage
leverages the temporal alignment of survey responses with digital data
collected via social media application programming interfaces (APIs) or web
scraping (Mellon, 2014; Stier et al., 2018). Due to the “site-centric” strategy
that focuses on collective (vs. individual) digital data, it removes the burden
of user consent for digital data collection. On the other hand,
individual-level linkage collects personal digital traces from survey
respondents through smartphones, platform APIs, or browser plugins (Al Baghal et al., 2019; Vaccari et al., 2015; Wells \&
Thorson, 2015). The “user-centric” process requires survey respondents to give
informed consent for the collection of their historical digital trace data.
This study adopts the “user-centric” approach, as it enables a more nuanced
examination of online interactions at the individual level, ensuring ecological
validity and ethical integrity in digital media research. 

Empirical research on data linkage
focuses on online news exposure, political knowledge and participation (Guess
et al., 2023; Kristensen et al., 2017; Song \& Cho, 2021; Thorson et al.,
2021), leveraging social media data from Facebook (Thorson et al., 2021),
Twitter (Freelon et al., 2024), and Instagram (Driel et al., 2022). So far,
there has been no research leveraging data linkage to understand selective engagement
with health-related TikTok videos.

\subsection{Impacts of Engagement with Health-Related TikTok Videos}

Survey research in health
communication and social media has established the link between self-reported
social media consumption and health risky behaviors among young adults,
including the use of e-cigarettes (Pokhrel et al., 2018; Yang et al., 2019,
2023) and alcohol (Geber et al., 2021; Hendriks et al., 2021). In terms of
social media engagement, young adults tend to report low levels of engagement
with tobacco-related social media content (Hébert et al., 2017), although anti-tobacco
engagement (e.g., sharing anti-tobacco messaging) is reported to be more common
than pro-tobacco engagement (Clendennen et al., 2020) on social media. Based on
the selectivity paradigm (Knobloch-Westerwick et al.,
2017), individuals information consumption and engagement choices can reinforce
existing beliefs and behaviors or prompt changes. Given the predominance of
pro-tobacco and pro-drinking content on TikTok (Russell et al., 2021; Sun et
al., 2023; Xie et al., 2023), engaging with pro-smoking and pro-drinking content
(vs. anti-smoking and anti-drinking) content is likely to foster positive
attitudes toward smoking and increase the frequency of smoking behaviors. Yet,
a longitudinal survey study shows that self-reported engagement with both pro-
and anti-tobacco on social media is positively associated with the use of
e-cigarettes (Yang et al., 2023). 

The inconclusive findings could be
attributed to the usage of self-reports to measure social media consumption and
engagement, which often represent an inaccurate estimate of actual media use (Araujo
et al., 2017; De Vreese \& Neijens, 2016; Parry et
al., 2021; Scharkow, 2019).  In
particular, when it comes to self-reporting the engagement with socially
undesirable behaviors (Song \& Cho, 2021), participants are likely to
underreport their engagement with smoking or drinking usually considered as
health risky behaviors. Digital traces offer a more objective and reliable
method to measure engagement with videos on smoking and drinking. They also
allow for a more ecologically valid and objective quantification of selective
engagement by calculating the ratio of such content to all content liked by an
individual at different time points. Moreover, the convergence perspective
suggests that individuals selectively consume a collection of content (Zhao et
al., 2022), which could involve diverse health topics. As people process one
message in relation to another (Akin et al., 2019), it is crucial to examine
how people’s engagement with a mix of health topics affects their vaping and
drinking behaviors. To address these limitations, this study compares TikTok
engagement measured by digital traces of likes versus self-reports and employs
data linkage to examine the impact of selective TikTok engagement on
individuals’ self-reported vaping and drinking behaviors. 

\subsection{Summary of Hypotheses and Research Questions}

Based on the literature review on the selectivity paradigm, social media engagement with
health topics, and data linkage approach, we conduct a computational analysis
of health topics liked by individuals on TikTok and link the analyzed topics
with their self-reported behavioral outcomes. To increase generalizability, we focus
on two health risky behaviors: smoking and drinking. First, the reciprocal
influence of user interactive behavior and algorithmic filtering may amplify
exposure and engagement with certain types of content. As such, TikTok users
who liked smoking or drinking videos initially may find themselves liking more
of such content over time. In other words, users tend to like more TikTok
videos over time, and this trend could be either constant or accelerating. We thus
ask the first research question:

\textbf{RQ1:} What is the temporal pattern of TikTok
users liking videos containing smoking (a) and drinking (b) content? Does the
trend of liking such videos stay at a constant or an accelerating rate? 

Based on the selectivity paradigm (Valkenburg,
2022), individuals’ information engagement choices can reinforce existing
behaviors or prompt changes. Thus, the size of the liked videos on smoking and
drinking should be positively associated with self-reported vaping and drinking
behaviors. 

\textbf{H1:} Controlling for demographic factors, the size of liked
videos containing smoking (a) and drinking (b) content are associated with
self-reported vaping (a) and drinking (b) behaviors. 

Individuals can encounter a
collection of TikTok videos on diverse health topics in their personalized
information environment. As people process one message in relation to another (Akin et al., 2019), it is crucial to examine how people’s engagement with a mix of
health-risk topics affects their vaping and drinking behaviors. Based on the
literature on health-risk communication, marijuana use is a behavioral risk
factor associated with vaping and drinking, while vegetable and fruit
consumption and exercising are behavioral factors reducing risks of vaping and
drinking (e.g., Uddin et al., 2020; Wilson et al., 2022). The following
research question is thus asked: 

\textbf{RQ2:} How is the size of liked videos containing additional health
topics (i.e., vegetable and fruit consumption, exercise, marijuana) associated
with self-reported vaping and drinking behaviors? 

The literature suggests that in
general, individuals report low engagement with social media content on smoking
and drinking, and the self-reported engagement with anti-smoking or
anti-drinking content is higher than pro-smoking or pro-drinking content. As content
of different valence might affect actual behaviors in different ways, we ask
the following research question:

\textbf{RQ3:} What is the actual prevalence of
smoking and drinking related TikTok videos of different valence (positive vs. non-positive)? How does the actual valence of such videos affect self-reported behavioral outcomes?

Last, we ask a research question to understand
the implications of measuring TikTok engagement using digital traces versus
self-reports. As digital traces can be collected at different time points and might
fluctuate, we aim to explore how these temporally fluctuating traces correspond
with self-reported measures and actual behaviors by comparing objective digital
traces of TikTok liking at different points and self-reported engagement and
behavioral measures.

\textbf{RQ4:} How do digital engagement traces at different time segments correlate with self-reported engagement and behavioral
measures?

\section{Method}

\subsection{Data Collection and Linkage}

Following IRB approval, data were collected through an online panel administrated by Qualtrics
from July to October 2023. Participants were U.S. residents aged 18 and older
who had TikTok accounts. They were asked to complete the questionnaire using a
desktop, laptop, tablet, or iPad. This requirement was set because a separate
mobile device was necessary for them to grant authorization for account access,
facilitating data linkage. Participants first completed a survey about their
TikTok use, beliefs, attitudes, and behaviors related to vaping, drinking, and
binge drinking, along with demographic questions. Then, participants had the
option to consent to share their TikTok data via a third-party application,
TikAPI, with a small incentive offered for this additional step. Those
consenting logged into their TikTok accounts on their mobile devices, agreed to
TikAPI’s terms, and authorized data scraping. Participants’ login credentials
were securely stored in TikAPI’s database, making them inaccessible to
researchers. Participants were later provided instructions via email on how to
manage and delete their data from TikAPI, ensuring their privacy and control
over their information.

Quota sampling was used to intentionally oversample racial/ethnic minorities,
ensuring an equitable representation of diverse groups in the sample. A total
of 1,102 people participated in this study, with 166 participants (15\%) agreeing
to authorize us access
to their TikTok profiles and content through TikAPI (2023). Our sample for this
study only included the 166 participants who provided both digital traces and
survey self-reports. In the sample, the average number of liked videos was
2,409 (\textit{Median} = 698, \textit{SD }= 2,919, \textit{min} = 0, \textit{max} =
7,709). A majority of participants were aged between
21 to 30 years old (33.73\%) and between 31 to 40 years old (22.29\%). The
distribution of other age groups was as follows: 20 or below (12.05\%), 41 to 50
years (14.46\%), 51 to 60 years (10.24\%), and 61 or above (7.22\%). In terms of sex,
69.88\% identified as female, 27.11\% identified as male, and 3.01\% identified as
non-binary. In terms of education, most participants held bachelor’s or
associate degrees (32.53\%), had some college education without a degree
(31.33\%), or were high school graduates (22.29\%). A smaller percentage had a master’s
degree or higher (10.84\%), while a minimal fraction (3.01\%) had less than a
high school degree. For race/ethnicity, 46.39\% identified as White/Caucasian,
21.08\% Hispanic/Latino, 18.67\% Black/African American, and 13.86\% Asian,
Pacific Islander, American Indian, or Native American. Despite the smaller
sample size, our sample captured a diverse group of TikTok users.

To manage the extensive number of
videos liked by participants and address TikAPI’s limitation of 2,000 API
requests per day, we created a comprehensive keyword list for health-risk
topics (e.g., e-cig, liquor, tobacco, addiction, cancer) (see Appendix 1). Our
focus on health-risk topics in the sampling scheme was justified by the
project’s focus. The average number of health-related topics liked by
participants was 131.30 (\textit{Median} = 29.00, \textit{SD} = 181.50, \textit{min }=
0, \textit{max} = 948). As such, we randomly sampled and download up to 200 health-related
videos for each participant for subsequent analysis. Specifically, if a
participant had liked more than 200 videos, only 200 were downloaded; if they
had liked fewer than 200 videos, all were downloaded. This approach allowed us
to download videos and user profiles for 5--10 participants per day, given the
API request constraints. As a result, a total of 13,724
videos were downloaded using a Python script, amounting to 93.01 GB of data.

\subsection{Computational Analysis of Liked TikTok Videos}

The computational analysis of TikTok
videos aims to convert as much audiovisual information as possible into
structured data including continuous or categorical variables. For the audio
track of each video, we used a speech-to-text neural engine from Deepgram, Inc.
(Deepgram, 2023) to transcribe speakers’ sentences into text form. The neural
engine additionally provided the topics estimated from the transcription and
encoded them in common such terms as “drinks,” “tobacco,” “fruits,”
“vegetables,” and “exercise.” Each estimated topic is also associated with a
confidence level from the neural engine. Since the unit of analysis is at the
participant level, we aggregated for each participant the topics and confidence
levels from all their liked videos. The sum of all confidence levels was
recorded to reflect the overall strength of topics associated with a
participant, and each topic’s confidence level was normalized by dividing the
sum of confidence levels to allow comparison across participants (i.e., size of
liked videos on smoking or drinking).

We also repeated the above steps to
separately generate the size of liked videos for the following time segments:
the recent one year (09/2022--09/2023),
the year before the recent year (09/2021--09/2022),
and the earlier period (04/2020--09/2021).
Note that TikAPI’s data access limitations prevent retrieving liked videos from
before 2020. This aims to answer RQ4, which explores the correlations between
digital traces of TikTok liking at different time points and self-reported
measures.
We also explored using state-of-the-art video summarization deep
learning algorithms (e.g., Iashin \& 
Rahtu, 2020) to summarize the visual component of the
TikTok videos. However, these algorithms produced only high-level summaries.
For example, the algorithm by Iashin and 
Rahtu (2020) gave such descriptions as “A man is sitting on
a chair” and “A man is standing in a kitchen with a woman in a white shirt,”
likely due to a misalignment in training tasks compared to the specific task of
describing TikTok video content. 
Hence, video summaries generated by the
current generation of deep learning algorithms are far less specific and useful
than audio transcripts. Considering 74.8\% of our TikTok videos contain human speech content that was successfully transcribed, we decided to utilize solely the audio information for computational analysis.

For the trend analysis of liked
drinking videos, we retrieved videos with transcriptions containing keywords
“alcohol,” “beer,” “drink,” or “wine” and recorded their respective posting
time as a proxy for participants viewing time. We further grouped videos into
time buckets of one month for subsequent time series analysis. Using posting
time as a proxy for viewing time combined with time-binning results in a
reasonable approximation as TikTok videos are typically viewed within a short
period after being posted. Trends analysis was conducted for each participant
individually on the time series of cumulative count (i.e., number of liked
drinking videos) as a function of the month index. An \textit{F}-test was
conducted between the null hypothesis of a linear line (i.e., viewing counts
evenly spread over time; cumulative count $\sim$ time + intercept) versus the
alternative hypothesis of a quadratic line (viewing counts linearly increase or
decrease over time; cumulative count $\sim$ time + time$^2$ + intercept). 

Further, for drinking related videos, we coded the transcripts’ valence (positive = 1 or nonpositive = $-$1) toward drinking using OpenAI’s GPT-4 API (for the prompt, see Appendix 2). GPT
has been shown in the literature to perform coding tasks comparable to human
coders in social sciences (Ziems et al., 2024). For each participant, the
valence of their liked TikTok videos was measured by both the number of
positive (\textit{M} = 5.14, \textit{SD} = 10.00) and nonpositive videos (\textit{M}
= 2.54, \textit{SD} = 3.56).

\subsection{Measures of Self-Reported Variables}

\textbf{Former use of e-cigarettes.} Respondents indicated whether they have ever tried an e-cigarette or vaping device, even once or twice (Pearson et al., 2018). 61.4\% indicated yes, whereas 38.6\% indicated no.

\textbf{Current frequency of e-cigarette use.} If a respondent
had previously used an e-cigarette or engaged in vaping, they were asked to
report their current frequency of using an e-cigarette or vaping device
(Pearson et al., 2018). This was measured on a 5-point scale from 1 \textit{(not at
all}) to 3 (\textit{less than weekly, but at least once a month}) to 5 (\textit{daily
or almost daily}). The mean of current vaping frequency was 3.02 (\textit{SD}
= 1.51). 

\textbf{Current frequency of alcohol use.} On a 6-point scale from 1 (\textit{never}) to 3 (\textit{1-3 days a month}) to 6 (\textit{5 or 7 days a week)}, respondents
indicated their current frequency of having any kind of drink containing
alcohol (\textit{M} = 3.02, \textit{SD} = 1.51), as well as the frequency of having
five drinks or more on one occasion (\textit{M} = 2.18, \textit{SD} = 1.45) (Yang
\& Zhao, 2018). 

\textbf{Self-reported engagement with vaping TikTok videos.} Following
existing measures (Hébert et al., 2017), respondents reported the
frequency that they sought information on e-cigarettes or vapes on TikTok and
they talked to other TikTok users about e-cigarettes or vapes on a 7-point
scale from 1 (\textit{not at all}) to 7 (\textit{a lot}). The mean of self-reported
engagement was 1.90 (\textit{SD} = 1.52, Pearson’s \textit{r} = 0.78). On the same
scale, respondents also reported the extent to which they had paid attention to
TikTok content on e-cigarettes or vaping devices (\textit{M} = 3.06, \textit{SD} =
1.74).

\textbf{Control variables.} Demographic
variables including age group,
sex, race/ethnicity, and education were controlled. To control the level of
social media activity and engagement, the total number of following and total
number of hearts (i.e., likes on TikTok) was also used as a covariate. 
Former use of e-cigarettes was also used as a covariate.

\section{Results}

RQ1 examined the trend in the rate of liking smoking and drinking related TikTok videos
over time. Due to the limited number of participants liking smoking-related
videos, the analysis primarily concentrated on the trend associated with
drinking videos. Among 92 participants who viewed any drinking-related videos,
35 viewed more than 4 videos across different years, enabling visualization and
\textit{F}-tests on the trend in the rate of liking drinking videos. Among them,
12 (34.29\%) participants demonstrated a statistically significant accelerating rate
of liking drinking TikTok videos (\textit{p} $<$ .05), whereas 23 demonstrated a
constant rate of liking such videos. We visualized the distinct trends by
setting time on the horizontal axis and the cumulative count of relevant TikTok
videos on the vertical axis. Figure 1 shows the significant trend (i.e., an
increasing rate of liking drinking videos) for 12 participants over time, whereas
Figure 2 shows the nonsignificant trend (i.e., a constant rate of liking). Our
analysis also shows that the presence of an accelerating rate was not related
to the total number of liked drinking videos.

\begin{figure}[!t]
\centering
\includegraphics[width=0.6\linewidth]{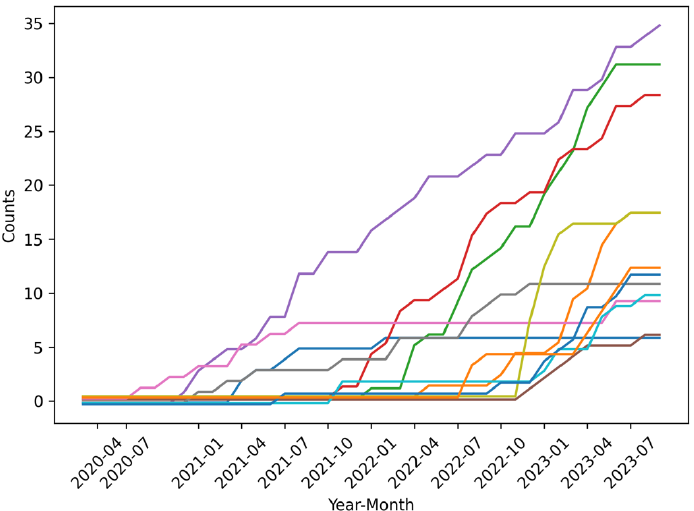}
\caption{Significant trend of an increasing rate of liking drinking TikTok videos.}
\end{figure}

\begin{figure}[!t]
\centering
\includegraphics[width=0.6\linewidth]{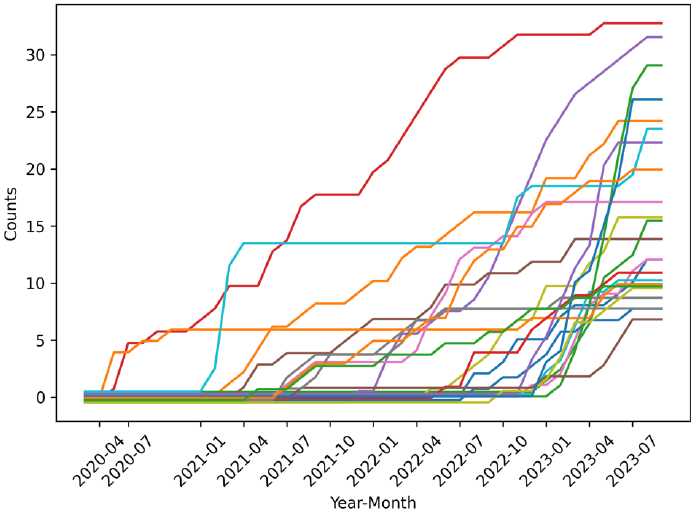}
\caption{Insignificant results: A constant rate of liking drinking TikTok videos.}
\end{figure}

H1 hypothesized that the size of liked
videos containing smoking and drinking was associated with self-reported vaping
and drinking behaviors, with demographic variables being controlled for. RQ2
asked how the size of liked videos containing additional health topics (vegetable
and fruit consumption, exercise, marijuana) was associated with self-reported
vaping and drinking behaviors. The results from linear regressions are in Table
1. We found that current self-reported drinking and binge drinking behaviors
were positively affected by the size of liked TikTok videos on drinking (\textit{b}
= 6.03, \textit{SD} = 2.83, \textit{p} = .046 for drinking; \textit{b} = 5.69, \textit{SD}
= 2.66, \textit{p} = .045 for binge drinking) and negatively affected by the size
of liked TikTok videos on fruit and vegetable consumption (\textit{b} = $-$8.86, \textit{SD}
= 4.26, \textit{p} = .042 for drinking; \textit{b} = $-$8.47, \textit{SD} = 4.00, \textit{p}
= .038 for binge drinking). The current self-reported vaping behaviors were
also positively affected by the size of liked TikTok videos on smoking (\textit{b}
= 2.90, \textit{SD} = 1.36, \textit{p} = .034). H1 was supported. For RQ2,
self-reported current drinking and vaping behaviors were only associated with
liking TikTok videos about vegetable and fruit consumption, not those about
physical exercise or marijuana. 

\begin{table}[!t]
\caption{Predictors of Self-Reported Drinking and Vaping Behaviors}
\includegraphics[width=1.0\linewidth]{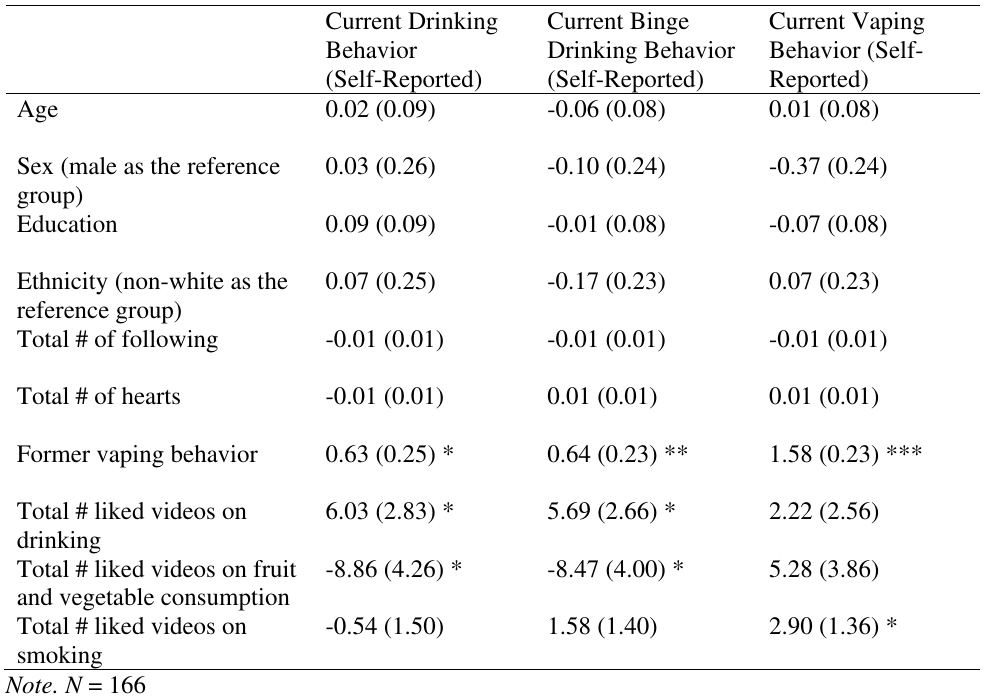}
\end{table}

RQ3 asked about the prevalence of smoking and drinking related
TikTok videos of different valence (positive vs. non-positive) and the effect
of valence on self-reported behaviors. We focused on drinking related videos due
to the small number of liked smoking videos. Our results showed that overall,
more participants liked pro-drinking (57.61\%) rather than anti-drinking TikTok
videos (42.39\%). But neither liking pro- or anti-drinking content related to
the self-reported current drinking or binge drinking behaviors. Additional
exploratory analyses show that liking pro-drinking content only negatively
related to affective outcomes, namely stress (\textit{b} = $-$0.02, \textit{SD} =
0.01, \textit{p} = 0.043) from TikTok use.

\begin{table}[!t]
\caption{Drinking-Related Liking Traces on TikTok and Self-Reported Drinking Behaviors}
\includegraphics[width=1.0\linewidth]{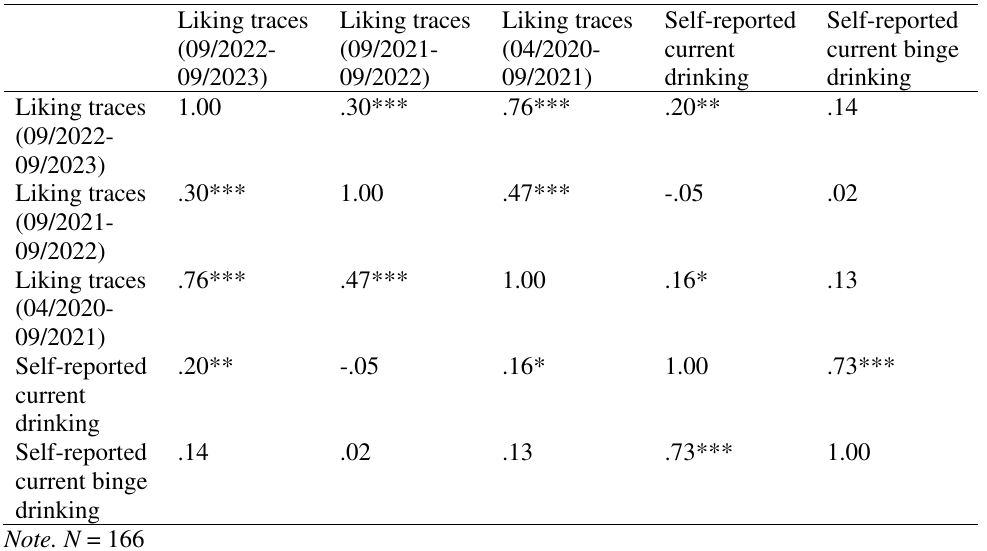}
\end{table}

\begin{table}[!t]
\caption{Smoking-Related Liking Traces on TikTok, Self-Reported Engagement Measures, and Vaping Behaviors}
\includegraphics[width=1.0\linewidth]{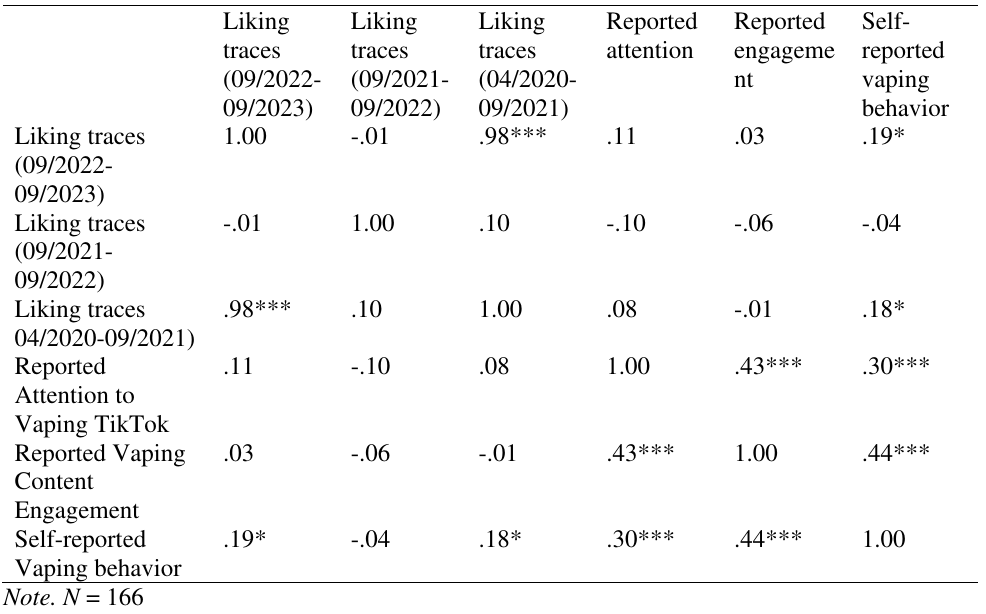}
\end{table}

RQ4 asked how digital engagement traces in
different time segments correlated with self-reported engagement and behavioral
measures. Our correlation tests showed that TikTok liking traces for drinking
and smoking in the recent one year (09/2022--09/2023) and in the earlier period (04/2020--09/2021)
strongly correlated
with each other (Tables 2 and 3). Liking traces in the year before the recent
year (09/2021--09/2022)
showed no correlation with the previous two traces for vaping and only
demonstrated moderate correlations for drinking. Additionally, liking traces at
different times did not correlate with self-reported attention to or engagement
with vaping-related content. And self-reported behaviors of drinking and vaping
had a slightly larger association with liking traces in the recent one year compared
to all in the past. However, these behaviors did not correlate with liking
traces in the mid-time segment.

\section{Discussion}

Our finding, based on the linkage of
longitudinal TikTok engagement traces with survey self-reports, reveals that
users who initially liked drinking-related videos tend to like more of such
content over the years. However, individual users display varied patterns: some
exhibit a linear trend, consistently liking an increasing number of drinking
videos over the years, while others show a nonlinear increasing trend,
indicating an accelerating rate of liking drinking videos on TikTok. This provides
empirical support for the literature on algorithmic impact (Thorson et al., 2021)
and selective engagement, which assumes that the reciprocal influence of algorithmic filtering and user engagement
behavior (e.g., liking) might affect or even amplify ones’ engagement with
certain types of content. This also suggests that TikTok’s algorithm might not
have a unified impact on all users, and additional factors such as individual
demographics, attitudes, or actual behaviors could account for the varying
rates at which users like such videos over time. Probably, some users might
interchangeably or complementarily use TikTok and other platforms to engage
with content related to drinking. Taken together, these findings expand the
selectivity paradigm and convergence theories by revealing temporal patterns of
selective engagement on an emerging platform. 

Additionally, we found that the actual size of liked TikTok videos on drinking and smoking,
extracted from all past activity traces through computational methods,
positively affected users’ self-reported current drinking, binge drinking, and
vaping behaviors. 
These findings extend existing studies in health communication and social media by
validating potential causal links between social media engagement over time and
health-risk behaviors (Pokhrel et al., 2018; Yang et al., 2019). This was
achieved by correlating actual engagement traces on TikTok with survey
self-reports. Our trace data also revealed that more participants liked TikTok
videos related to drinking rather than smoking or vaping. Among those who liked
drinking-related content, there was a slight preference for pro-drinking over
anti-drinking videos. However, the valence of the liked videos did not
influence actual drinking behaviors. Our exploratory analysis shows that the valence
of the liked videos related to affective outcomes, implying that indirect
relationships might exist between valence and behavioral outcomes. 
Additionally, we found that the actual size of liked videos on fruit and vegetable consumption, negatively affected all these
behaviors. This suggests that both social media engagement related to certain
risky behaviors and auxiliary engagement related to other health behaviors can
collectively affect behavioral outcomes. According to the convergence
perspective (Zhao et al., 2022), individuals can encounter a collection
of TikTok videos on diverse health topics in their personalized information environment.
As people process one message in relation to another (Akin et al., 2019), it is
crucial to examine the behavioral outcomes of the engagement with a mix of
health topics. This convergence perspective might provide a more ecologically valid standpoint for understanding the behavioral impact of social media engagement. 

Further, we found that TikTok liking traces on drinking and smoking in different time segments
strongly correlated with each other. Yet, these liking traces in different time
segments did not correlate with self-reported engagement with vaping-related
content. This supports the methodological literature that the usage of
self-reports to measure social media consumption or engagement can represent an
inaccurate estimate of actual media use (Araujo et al., 2017; Parry et al.,
2021; Scharkow, 2019). As drinking and smoking are
considered as socially undesirable behaviors, participants probably underreported
their engagement with smoking or drinking TikTok videos. Our digital traces offer
a more objective and reliable method to measure engagement with videos and the
subsequent impacts. Our observations that engagement traces in the recent one
year and all past years exhibited strong correlations suggest that digital
traces within a recent time frame might well represent all past traces in
predicting behavioral outcomes. However, engagement traces in the mid-time
segment had weaker correlations with those in other time segments, suggesting
the potential fluctuations of user engagement traces over time and the
importance of adopting a reasonable period in data collection.

\subsection{Methodological and Theoretical Implications}

Our results significantly enhance the
methodological literature, especially in the less explored areas of data
linkage approach and computational analysis of audiovisual content. The finding
that health-related topics, computationally extracted from TikTok videos liked
by users, did not align with their self-reported engagement measures
corroborates existing studies on the potential bias of self-reports (Araujo et
al., 2017; Naab et al., 2018), particularly in health risky behaviors. Digital
traces and computational measurement serve as promising methods to obtain
objective and reliable measures of social media consumption. The varying
correlations observed in engagement traces over different time periods
underscore the significance of data collection period. Selecting an appropriate
timeframe for data collection is crucial, as it allows scholars to detect the
hypothesized effects of social media engagement.

Emerging methodological literature
has revealed the potential of combining data linkage and computational methods (e.g.,
Freelon, 2024). While future research can apply this approach to study the
longitudinal impacts of social media platforms and their downstream behavioral
effects, complications in system design, social media API constraints/updates,
and growing user privacy concerns may continue to pose challenges. For example,
only a small percentage of participants (15\% in this study) were willing to provide
access to their historical TikTok data, potentially limiting the
generalizability of the results and making this method costly for most scholars.
To mitigate these challenges, increased efforts in dataset sharing within the
scholarly community and open-source sharing can enhance data accessibility and
method generalizability, thereby strengthening research robustness and
reproducibility of research findings. 

Theoretically, our research enriches
the selectivity paradigm in communication theories by explicating and testing the
concept of selective engagement on an emerging social media platform. Enriching
the selectivity paradigm with research on social media engagement, convergence
perspective, and algorithmic impact, our findings also show that engaging with
a mix of health-risk videos on TikTok can differentially affect health risky
behaviors. The combination of TikTok engagement traces and survey self-reports
enables an ecologically valid test of not only the longitudinal trends of
TikTok user engagement with health-related videos but also the behavioral impacts
of such engagement.

\subsection{Limitations}

Our findings should be generalized
to the broader population with caution. Despite controlling for demographic
variables, only 15\% of our survey sample consented to and successfully provided
access to their TikTok activity traces. These participants might possess higher
digital literacy and better understand algorithmic impact than average TikTok
users, potentially leading them to intentionally curate content that caters to
their preferences. 
Future studies should, if feasible, aim to obtain digital
traces from a more representative sample. Additionally, our study focused
solely on the digital traces of likes on TikTok, neglecting other forms of
engagement due to data unavailability. Future research should examine various
forms of selective engagement across multiple platforms. Last, although we have
individuals’ TikTok traces from 2020 to 2023, we lack longitudinal behavioral
measures, which limits our ability to make causal conclusions. Future research
can combine longitudinal self-reports with digital traces to understand the
reciprocal influence of selective engagement on social media and behavioral
outcomes. 

\section{Conclusion}

Despite its limitations, our study emphasizes the importance of integrating digital traces,
computational analysis, and self-reported data for an objective and
ecologically valid approach to studying media consumption and engagement on digital
platforms. This methodological innovation significantly enhances our ability to
accurately assess social media’s impact on behavior and opens opportunities for
theoretical building in communication.

\newpage
\begin{center}
  {\LARGE Online Supplemental Materials}\vspace{4mm}\\
  {\large TikTok Engagement Traces Over Time and Health Risky Behaviors: Combining Data Linkage and Computational Methods}
\end{center}

\appendix
\renewcommand\thesection{Appendix~\arabic{section}}
\section{Sampling Keywords}
vape, vaping, vaper, smoke, smoking, smoker, jull, \#pod, \#nic, inhaler, e-juice, \#vapelife, \#vapor, e-liquid, e-cig, e-cigarette, cigarette, ecigarette, cig, e-cigg, e-ciig, \#ecigg, juice, liquid, waterpipe, cigar, weed, 
marijunna, fumar, cigarro, \#quitvaping, \#quittingnicotine, \#nicotinefree, \#vapetricks, \#nicotinefein, nic, \#iget, puff bar, "\#puffbar", exhale, inhale, vapor, addicted, vgod, blowos, vemjuice, tobaco, tobacco, quit, nicotine, lungs, lungcancer, cancer, \#disposable, nictok, \#quittingnic, drink, alcohol, beer, wine, drunk, drank,sober, \#binge, \#titos, liquor, drinktok, vodka, cocktail, 
addiction, \#sobertok, hangover, bar, bartender

\section{GPT 4 Prompts}

\begin{lstlisting}
You are a social scientist good at content analysis and coding. You will be given a set of transcripts in JSON array format. Please analyze for each transcript the stance (i.e., positive=1 or non-positive=-1) toward drinking. If the transcript is not related to drinking but is present, please output 0. Please output the stance coding in a list in compact JSON format. Do not explain, and only give the final result.
\end{lstlisting}

\end{document}